\documentclass[a4paper,preprint,showpacs,amssymb,superscriptaddress]{revtex4}
\usepackage{graphicx}

\begin{document}

\title{Nonequilibrium Free Energy-Like Functional for the KPZ Equation}

\author{Horacio S. Wio}

\affiliation{Instituto de F\'{\i}sica de Cantabria (IFCA),
Universidad de Cantabria \& CSIC, E-39005 Santander, Spain}

\begin{abstract}
Opposing to a (common) belief against the existence of a
thermodynamic-like potential for the KPZ equation, here we present a
derivation for such a functional. With its knowledge we prove some
global shift invariance properties previously conjectured by other
authors. The procedure could be extended in order to derive a more
general form of such a functional leading to other known related
nonlinear kinetic equations. Exploiting the KPZ's functional, and
for arbitrary dimension, we have obtained the exact form of the
stationary probability distribution function and have shown a couple
of examples of how it is possible to exploit it in order to obtain
relevant results like finding support to the conjecture that in the
strong coupling regime a critical dimension doesn't exists.
\end{abstract}


\maketitle

\normalsize

Phenomena far from equilibrium are ubiquitous in nature, including
among many other, turbulence in fluids, interface and growth
problems, chemical reactions, biological systems, as well as
economical and sociological structures. During the last decades the
focus on statistical physics research has shifted towards the study
of such systems. Among those studies, the understanding of growing
kinetics at a microscopic as well as on a mesoscopic level
constitutes a major challenge in physics and material science
\cite{x1,x2,x3,x4}. Some recent papers have shown how the methods
and know-how from static critical phenomena have been exploited
within nonequilibrium phenomena of growing interfaces, obtaining
scaling properties, symmetries, morphology of pattern formation in a
driven state, etc
\cite{hents,spohn-1,fogedby-1,juanma,ma-yang,castro}.

It is a common belief that the nontrivial spatial-temporal behavior
occurring in several nonequilibrium systems \cite{rev1}, originates
from the {\it non-potential} (or \textit{non-variational}) character
of the dynamics, meaning that there is no Lyapunov functional for
the dynamics. However, Graham and co-workers have shown in a series
of papers \cite{GR1} that a Lyapunov-like functional exists for very
general dynamical systems, like the complex Ginzburg-Landau
equation. Such a functional is formally defined as the solution of a
Hamilton-Jacobi-like equation, or obtained in a small gradient
expansion \cite{GR1,GR2}. The confusion associated with the
qualification of \textit{nonvariational} dynamics comes from the
idea that the dynamics of systems having nontrivial attractors
(limit cycle, chaotic) cannot be deduced from the minimization of a
potential playing the same role as the free energy in equilibrium
systems \cite{maxi1,me1,me2}. Nevertheless, this does not preclude
the existence of a Lyapunov functional for the dynamics that will
have local minima identifying the attractors of the system. However,
once the system has reached an attractor that is not a fixed point,
the dynamics proceeds inside the attractor driven by
\textit{nonvariational} contributions to the dynamical flow, that do
not change the value of the Lyapunov functional, implying that the
dynamics is not completely determined once the indicated functional
is known. This situation has some known examples even in equilibrium
statistical mechanics \cite{HH}. Hence, the Lyapunov functional, or
\textit{nonequilibrium potential} (NEP) \cite{GR1}, plays the role
in nonequilibrium situations of a thermodynamic-like potential
characterizing the global properties of the dynamics: attractors,
relative (or nonlinear) stability of the attractors, height of the
barriers separating attractions basins, and offers the possibility
of studying transitions among the attractors due to the effect of
(thermal) fluctuations.

In a recent series of papers we have shown several results related
to the obtention and exploitation of the indicated NEP's concept in
scalar and non-scalar reaction-diffusion systems (see \cite{me2} and
references therein). In particular we have exploited those results
for the study of stochastic resonance \cite{SR} in extended systems
(see \cite{me1,me2,nos3} and references therein). In those works, we
have analyzed problems of stochastic resonance in scalar and
activator-inhibitor systems, systems with local and nonlocal
interactions, system-size stochastic resonance, etc.

Here, and related to the kinetics of growing interfaces, we discuss
the case of the Kardar-Parisi-Zhang equation (KPZ) \cite{kpz}. This
equation describes the evolution of a field $h(\bar{x},t)$, that
corresponds to the height of a fluctuating interface,
\begin{eqnarray}\label{eq-000}
\frac{\partial}{\partial t} h(\bar{x},t) = \nu \nabla^2 h(\bar{x},t)
+ \frac{\lambda}{2} \left( \nabla h(\bar{x},t) \right)^2 + K_o + \xi
(\bar{x},t),
\end{eqnarray}
where $\xi (\bar{x},t)$ is a Gaussian white noise, of zero mean
($\langle \xi (\bar{x},t) \rangle = 0$) and correlation $\langle \xi
(\bar{x},t) \xi (\bar{x}',t') \rangle = 2 \varepsilon \delta
(\bar{x}-\bar{x}') \delta (t-t')$. As indicated above, this
nonlinear differential equation describes fluctuations of a growing
interface with a surface tension given by $\nu$, $\lambda$ is
proportional to the average growth velocity and arises because the
surface slope is parallel transported in such a growth process.

Opposing to a claim in a recent paper \cite{fogedby-1}: \textit{The
KPZ equation is in fact a genuine kinetic equation describing a
nonequilibrium process in the sense that the drift $\nu \nabla^2 h +
\frac{\lambda}{2} (\nabla h )^2 - F$ cannot be derived from an
effective free energy}; we show here that such a nonequilibrium
thermodynamic-like functional (NETLP) for the KPZ equation exists.
Exploiting its knowledge, we will discuss conjectures advanced in
\cite{hents} and how they are fulfilled. We also briefly discuss how
to extent the derivation procedure in order to consider more general
forms of kinetic equations. Finally, we obtain the stationary (or
asymptotic) probability distribution function (pdf) --\textit{valid
for \textbf{any dimension} and unknown till now}-- and also derive
the form of the NEP and, through the analysis of a couple of simple
examples, we discuss how the knowledge of this NETLP and pdf could
be exploited in order to obtain some relevant results.

The Lyapunov functional or NETLP for the KPZ equation is given by
\begin{eqnarray}\label{eq-001}
{\cal F}[h] = \int_{\Omega} e^{\frac{\lambda}{\nu} h(\bar{x},t)}\,
\frac{\lambda}{4 \nu}\, \Bigl[ - K_o + \frac{\lambda }{2} \left(
\nabla h(\bar{x},t) \right)^2 \Bigr] d\bar{x}.
\end{eqnarray}
It is easy to prove that this functional fulfills both, the relation
\begin{eqnarray}\label{eq-002}
\frac{\partial}{\partial t} h(\bar{x},t) & = & - \Gamma [h]
\frac{\delta {\cal F}[h]}{\delta h(\bar{x},t)} + \xi (\bar{x},t),
\end{eqnarray}
as well as the (Lyapunov) property $\frac{\partial}{\partial t}{\cal
F}[h] = - \Gamma [h] \left( \frac{\delta}{\delta h}{\cal
F}[h]\right)^2 \leq 0$, where the function $\Gamma [h]$ is given by
\begin{eqnarray}\label{eq-002p}
\Gamma [h] = \left(\frac{2 \nu}{\lambda} \right)^2 e^{-
\frac{\lambda}{\nu} h(\bar{x},t)}.
\end{eqnarray}
Hence, from this \textit{free energy}-like functional, and by a
functional derivative, we can obtain the KPZ kinetic equation. It
corresponds to a relaxation model, analogous to model \textbf{A}
according to the classification in Hohenberg \& Halperin's review
\cite{HH}.

It is worth to make here a remark. In one hand, in the standard
``model A" it is known that the dynamics can be seen as a
superposition of modes that decay exponentially towards a steady
state, with time-dependent correlations obeying some constraints
such as positivity. But on the other hand, in the KPZ problem it is
known that the relaxation of perturbations decay following a
stretched exponential form \cite{z1,z2,z3,z4}.

In order to show how to obtain the above indicated functional, we
start considering the following simple scalar reaction-diffusion
equation for a positive ($\phi \geq 0$) field $\phi (\bar{x},t)$, as
it corresponds to a probability density,
\begin{equation}\label{eq-01}
\frac{\partial}{\partial t} \phi (\bar{x},t) = \nu \nabla^2 \phi
(\bar{x},t) + a \, \phi (\bar{x},t) + \eta (\bar{x},t) \, \phi
(\bar{x},t),
\end{equation}
where $a$ is a constant, and $\eta (\bar{x},t)$ is also a Gaussian
white noise of zero mean, and intensity $\sigma$, and we assume the
Stratonovich interpretation. It is well known that the system in Eq.
(\ref{eq-01}) has the following NETLP \cite{exact}
\begin{eqnarray}\label{eq-02}
{\cal F}_{o}[\phi] = \int_{\Omega} \left\{ - \frac{a}{2} \, \phi
(\bar{x},t)^2 + \frac{\nu}{2} \left( \nabla \phi (\bar{x},t)
\right)^2 \, \right\} d\bar{x},
\end{eqnarray}
where $\Omega$ indicates the integration range. As has been shown in
previous works \cite{nos1}, in addition to fulfilling the Lyapunov
propertiy $\frac{\partial}{\partial t}{\cal F}[\phi] \leq 0$, it
also fulfills the relation
\begin{eqnarray}\label{eq-03}
\frac{\partial}{\partial t} \phi (\bar{x},t) = - \frac{\delta {\cal
F}_o[\phi]}{\delta \phi (\bar{x},t)} + \phi (\bar{x},t) \, \eta
(\bar{x},t);
\end{eqnarray}
where the contribution from the boundaries is null, due to the
variation $\delta\phi$ being fixed there ($=0$), as usual.

Let us now define a new field, $h(\bar{x},t)$, that as indicated
before corresponds to the interface height, exploiting the so called
Hopf-Cole transformation $h(\bar{x},t) = \frac{2 \nu}{\lambda}\ln
\phi (\bar{x},t)$, with the inverse $\phi (\bar{x},t) =
e^{\frac{\lambda}{2 \nu} h(\bar{x},t)}.$ It is straightforward to
show that exploiting this transformation, the original Eq.
(\ref{eq-01}) becomes Eq. (\ref{eq-000}), with $a = K_o
\frac{\lambda}{2 \nu}$ and $\sigma = \frac{\lambda}{2 \nu}
\varepsilon.$ However, the noise term that in Eq. (\ref{eq-01}) has
a multiplicative character, in the transformed Eq. (\ref{eq-000})
becomes additive.

If we now apply the same transformation to the NETLP indicated in
Eq.(\ref{eq-02}), it is immediate to obtain the functional shown in
Eq. (\ref{eq-001}). Hence we have a \textbf{free energy-like
functional} from where the KPZ kinetic equation can be obtained
through functional derivation. Clearly, the contribution to the
variation that come from the boundaries is again null.

It is worth to consider once more Eq. (\ref{eq-01}), but now
including a typical limiting term of the form: $ - b \, \phi
(\bar{x},t)^3$. The resulting reaction-diffusion equation
corresponds to a version of the so called Schl\"{o}gl model
\cite{schlogl}. The associated NETLP will have an extra term of the
form $ + \frac{b}{4} \, \int_{\Omega} \phi (\bar{x},t)^4 \,
d\bar{x}.$ Applying once more the previously indicated Hopf-Cole
transformation, in Eq. (\ref{eq-000}) a new associated term arises,
having the form $ - \gamma \, e^{\frac{\lambda}{\nu} h(\bar{x},t)}$
($b = \gamma \frac{\lambda}{2 \nu}$). The new equation corresponds
to a form of the so called \textit{bounded-KPZ}
\cite{bound1,bound2}. Clearly, we will also have an extra term in
the associated NETLP (Eq. (\ref{eq-001})). However, in what follows
we consider the case $b = 0$, analyzing only the more ``usual" form
of the KPZ equation indicated by Eq. (\ref{eq-000}).

Let us now check some of the properties previously assumed for such
a functional. According to the analysis of global shift invariance
in \cite{hents}, it is easy to see that the relations indicated by
Eq. (9) in \cite{hents} are fulfilled. That is, we can readily prove
that if $l$ is an arbitrary (constant) shift
\begin{eqnarray}\label{eq-08}
{\cal F}[h + l] = K[l] {\cal F}[h]; \,\,\,\,\,\,\,\,\,\,\,\, \Gamma
[h + l] = K[l]^{-1} \Gamma [h],
\end{eqnarray}
with $ K[l] = e^{\frac{\lambda}{\nu} l} (\sim \Gamma [l]^{-1})$.

To prove other conjectures also indicated in \cite{hents}, we
introduce the  \textit{free energy-like density} $\widetilde{{\cal
F}}[h,\nabla h]$, which is defined by ${\cal F}[h] = \int d \bar{x}
\, \widetilde{{\cal F}}[h,\nabla h].$ The relations we refer are
\begin{eqnarray}\label{eq-08p}
\widetilde{{\cal F}}[h, \nabla h]= e^{sh} \widetilde{{\cal F}}_1
[(\nabla h)^2]; \,\,\,\,\,\,\,\,\,\,\,\, \Gamma [h, \nabla h]=
e^{sh} \Gamma _1 [(\nabla h)^2].
\end{eqnarray}
According to the form of the NETLP indicated in Eq. (\ref{eq-001}),
and the definition of $\widetilde{{\cal F}}[h,\nabla h]$, it is
clear that the first relation above results ``trivially" true. For
the second relation we have that $\Gamma [h, \nabla h] = e^{- s
h(\bar{x},t)} \Gamma_o$, where $\Gamma_o=1$, and
$s=\frac{\lambda}{\nu}$, as $\Gamma [h]$ is not function of $\nabla
h$. In addition, it can be also proved that the indicated NETLP is
invariant under the nonlinear Galilei transformation that, as
discussed in \cite{fogedby-1}, are fulfilled by the KPZ equation.

We can go still further and look for the possibility of deriving the
NETLP for other forms of related kinetic equations. To do that, we
should assume that we have a non-local reaction-diffusion equation,
as in \cite{me1},
\begin{eqnarray}\label{eq-21}
\frac{\partial}{\partial t} \phi (\bar{x},t) = \nu \nabla^2 \phi
(\bar{x},t)+ \, a \, \phi (\bar{x},t) - \beta \int_{\Omega}
d\bar{x'} \mathbf{G}(\bar{x},\bar{x'}) \phi (\bar{x'},t) +  \phi
(\bar{x'},t) \, \eta (\bar{x'},t),
\end{eqnarray}
where, as discussed in \cite{nos3} the kernel $\mathbf{G}(\bar{x},
\bar{x'})$ could be of a very general character, and $\beta$ is the
interaction intensity.

As we have done before, using the Hopf-Cole transformation we could
obtain a \textit{generalized} (\textit{nonlocal}) form of the KPZ
equation. However, in order to obtain sensible results we should
assume that the translational invariant kernel ($\mathbf{G}(\bar{x},
\bar{x'}) = \mathbf{G}(\bar{x} - \bar{x'})$) is of a very short
range and has some symmetry properties. What results is a nonlocal
contribution that, even differing from the one discussed in
\cite{nonloc1,nonloc2}, is of much interest and relevance. Repeating
the previous procedure we find the associated NETLP, but we will not
discuss this aspect here (see \cite{meLAWNP}). We only want to
remark that the contributions that arise have the same form of those
ones that arose in several previous works, where scaling properties,
symmetry arguments, etc, have been used to discuss the possible
contributions to a general form of the kinetic equation
\cite{hents,juanma,castro,argum2}. The different contributions that
arise are tightly related to several of other previously studied
equations, like the Kuramoto-Sivashinsky \cite{shiva}, the
Sun-Guo-Grant (SGG) equation \cite{sgg}, and others as in
\cite{hents} and \cite{castro}.

The knowledge of the NETLP for the KPZ equation allows us to readily
write the asymptotic long time probability distribution function
(pdf), valid for \textbf{any dimension}, which (due to the
``diagonal" character of $\Gamma [h]$) is given by
\begin{eqnarray}\label{eq-30}
{\cal P}_{as}[h(\bar{x},t)] & \sim & \exp \left\{ - \frac{2}
{\varepsilon} \int d\bar{x} \, \int ^{h(\bar{x},t)} _{h_r} d\psi \,
\Gamma [\psi] \frac{\delta {\cal F}[\psi]}{\delta \psi} \right\}
\nonumber
\\ & \sim & \exp \left\{  - \frac{2} {\varepsilon} \int d\bar{x} \,
\int ^{h(\bar{x},t)} _{h_r} d\psi \, \left( \nu \nabla^2 \psi
(\bar{x},t) + \frac{\lambda}{2} \left( \nabla \psi (\bar{x},t)
\right)^2 \right) \right\} \nonumber \\ & \sim & \exp \left\{ -
\frac{2}{\varepsilon} \int d\bar{x} \, \Bigl[ \Gamma [h]
\widetilde{{\cal F}}[h] - \int ^{h(\bar{x},t)} _{h_r} d \psi \,
\frac{\delta \Gamma [\psi]}{\delta \psi}
\widetilde{{\cal F}} [\psi] \Bigr] \right\} \nonumber \\
& \sim & \exp \left\{ - \frac{\nu}{2 \varepsilon} \int d\bar{x}
\left(\nabla h \right)^2 + \frac{\lambda}{2 \varepsilon} \int
d\bar{x}
\int ^{h(\bar{x},t)} _{h_r} d\psi \left(\nabla \psi \right)^2 \right\} \nonumber \\
& \sim & \exp \left\{- \frac{\Phi [h]}{\varepsilon} \right\},
\end{eqnarray}
where, in order to simplify, we assume $K_o = 0$, and $h_r$ is an
arbitrary reference state. The second line results from just
replacing the relation in Eq. (\ref{eq-002}), while the third and
fourth lines results transforming the pdf for the original field, or
by using functional methods. The first part of the last line shows a
nice structure, where we can identify a contribution, Gaussian on
the slope, plus a ``correction" proportional to $\lambda .$ For the
one-dimensional case, it is possible to show that only the well
known Gaussian result survives \cite{x2,x3}, if the adequate
boundary conditions are taken into account. Otherwise, the indicated
expression is the \textbf{complete} solution for the pdf,
irrespective of the boundary conditions.

The last line gives us the definition for $\Phi [h]$, that is
\begin{eqnarray} \label{nep001}
\Phi [h] = \frac{\nu}{2} \int d\bar{x} \left(\nabla h \right)^2 - \,
\frac{\lambda}{2} \int d\bar{x} \int ^{h(\bar{x},t)} _{h_r} d \psi
\, \left(\nabla \psi \right)^2,
\end{eqnarray}
a functional that could be identified as the \textit{nonequilibrium
potential} (NEP) \cite{GR1,GR2,me1} that fulfills
\begin{eqnarray}
\frac{\partial}{\partial \, t} h(\bar{x},t) = - \frac{\delta \Phi
[h]}{\delta h(\bar{x},t)} + \xi(\bar{x},t),
\end{eqnarray}
as well as $\frac{\partial}{\partial \, t}  \Phi [h] = -
\left(\frac{\delta}{\delta h(\bar{x},t)} \Phi [h] \right)^2.$ The
knowledge of $\Phi [h]$ opens the door to the exploitation of other
techniques \cite{GR1,GR2,me1}, offering alternatives to the usual
renormalization group ones. Clearly, it also shows that the claim
done in the phrase indicated at the beginning \cite{fogedby-1} is
not correct.

In order to show the possibilities that offers the knowledge of such
a NETLP, we now present a simple example. Let us consider a sightly
different case, where we only have a spatial quenched noise
(``disorder") instead of the spatio-temporal noise we have
considered so far. The associated equation is
\begin{eqnarray}\label{eq-100}
\frac{\partial}{\partial t} h(\bar{x},t) = \nu \nabla^2 h(\bar{x},t)
+ \frac{\lambda}{2} \left( \nabla h(\bar{x},t) \right)^2 + K_o +
\vartheta (\bar{x}),
\end{eqnarray}
where, as in previous studies \cite{disord1,disord2,disord3},
$\vartheta (\bar{x})$ is a quenched, Gaussian distributed, noise.
For this case we have that
\begin{eqnarray}\label{eq-101}
\frac{\partial}{\partial t} h(\bar{x},t) & = & - \Gamma [h]
\frac{\delta {\cal F}[h]}{\delta h(\bar{x},t)},
\end{eqnarray}
with the same form of ${\cal F}[h]$ as in Eq. (\ref{eq-001}), but
$K_o$ replaced by $K_o + \vartheta (\bar{x})$.

The indicated studies have shown that the front profile presents a
triangular structure \cite{disord3} and, clearly, it is of relevance
to determine the slope of such structures. Using the known form of
the NETLP, we can minimize this free-energy-like  functional and
obtain, in the 1-d case, that such a slope is $\alpha = \left(
\frac{16}{\nu \lambda}\right)^{1/3},$ a value that agrees quite well
with the numerical evaluations.

Summarizing, we have here found the form of the Lyapunov functional
or NETLP for the KPZ equation. From this NETLP, and through a
functional derivative, we have obtained the KPZ kinetic equation,
and have also shown that it fulfills global shift properties, as
well as other ones anticipated for such an unknown functional. Even
more, it is possible to extend the procedure to derive such a
functional, considering nonlocal terms, and in such a way derive
more general forms, that includes several other kinetic equations
studied in the literature of interface growing phenomena.

As indicated in the literature, dynamic renormalization group
techniques, being useful and powerful, in many cases only offers
incomplete results, having no access to the strong coupling phase
\cite{x2,wiese}. Hence, it is clear the need of alternative,
complementary, ways to analyze the KPZ and related problems (as an
example see \cite{sce,nonloc2} for a self-consistent expansion). The
present results clearly open new possibilities of making
non-perturbational studies for the KPZ problem. We expect that
through the analysis of long time mean values of $h(\bar{x},t)$, or
of correlations, we could extract information about scaling
exponents. Such study will be the subject of forthcoming work
\cite{nos6}.

\acknowledgments The author acknowledges the award of a {\it Marie
Curie Chair}, from the European Commission, during part of the
development of this work; and also thanks J.M. L\'opez, M.A.
Rodriguez, M.A. Mu\~{n}oz, J.A. Revelli and R. Toral for fruitful
discussions or comments.


\begin{thebibliography}{99}
\bibitem{exact} It is worth here remarking that the indicated form
of NETLP is exact, and not a small noise intensity approximation.
\end{thebibliography}
\end{document}